\documentclass[%
 reprint,
superscriptaddress,
 amsmath,amssymb,
 aps,
prb,
]{revtex4-1}

\usepackage{hyperref}

\hypersetup{}

\usepackage{units}
\usepackage[usenames, dvipsnames]{color}
\usepackage{graphicx}% Include figure files
\usepackage{dcolumn}% Align table columns on decimal point
\usepackage{bm}% bold math
\usepackage[english=nohyphenation]{hyphsubst}
\usepackage[english]{babel}
\usepackage{float}

\makeatletter
\newcommand*{\rom}[1]{\expandafter\@slowromancap\romannumeral #1@}
\makeatother

\begin{document}

%\preprint{APS/123-QED}

\title{Fast flux control of 3D transmon qubits using a magnetic hose}% Force line breaks with \\

\author{O. Gargiulo}
\affiliation{Institute for Quantum Optics and Quantum Information, A-6020 Innsbruck, Austria}
\affiliation{Institute for Experimental Physics, University of Innsbruck, A-6020 Innsbruck, Austria}

\author{S. Oleschko}
\affiliation{Institute for Quantum Optics and Quantum Information, A-6020 Innsbruck, Austria}
\affiliation{Institute for Experimental Physics, University of Innsbruck, A-6020 Innsbruck, Austria}

\author{J. Prat-Camps}
\affiliation{Institute for Quantum Optics and Quantum Information, A-6020 Innsbruck, Austria}
\affiliation{Institute for Theoretical Physics, University of Innsbruck, A-6020 Innsbruck, Austria}
\affiliation{{Interact Lab, School of Engineering and Informatics, University of Sussex, Brighton BN1 9RH, UK}}

\author{M. Zanner}
\affiliation{Institute for Quantum Optics and Quantum Information, A-6020 Innsbruck, Austria}
\affiliation{Institute for Experimental Physics, University of Innsbruck, A-6020 Innsbruck, Austria}

\author{G. Kirchmair}
\email{gerhard.kirchmair@uibk.ac.at}
\affiliation{Institute for Quantum Optics and Quantum Information, A-6020 Innsbruck, Austria}
\affiliation{Institute for Experimental Physics, University of Innsbruck, A-6020 Innsbruck, Austria}

\date{\today}% It is always \today, today,

\begin{abstract}

Fast magnetic flux control is a crucial ingredient for circuit quantum electrodynamics (cQED) systems. So far it has been a challenge to implement this technology with the high coherence 3D cQED architecture. In this paper we control the magnetic field inside a superconducting waveguide cavity using a magnetic hose, which allows fast flux control of 3D transmon qubits on time scales \unit[$< $100]{ns}. The hose is designed as an effective microwave filter to not compromise the energy relaxation time of the qubit. The magnetic hose is a promising tool for fast magnetic flux control in various platforms intended for quantum information processing and quantum optics.        %108 words

\end{abstract}

\pacs{Valid PACS appear here}% PACS, the Physics and Astronomy
\maketitle

%\textbf{Introduction} 
In many quantum systems used for quantum information processing, quantum optics and quantum enhanced measurements, magnetic fields are essential to mediate interactions or tune parameters.~\cite{taylor2008,chin2010,warring2013}  For individual quantum systems like qubits the typical wish-list is a combination of large field strength, the ability to change the field rapidly and apply it only locally while at the same time not introduce additional decoherence channels. Very often it is hard for experiments to simultaneously meet these criteria as e.g. fast switching times are limited by eddy or super-currents introduced by the metallic boundaries protecting the quantum system from decoherence.  %97 words

%NV centers: coupling and frequency control
%SC qubits: -"-
%Quantum dots: large fields locally applied

%no loss of coherence
%focus fields
%route over long distances

%fast timescale limited by eddy currents and/or SC currents

%- many quantum systems where fast b-field tuning is necessary

%- sc-qubits, quantum dots?, NV centers?, ...

%- hard to combine with experimental restrictions like eddy currents, sc, hard to focus b fields

%Need help here, especially refs.

%\textbf{Example SC qubits} 
One of the most promising platforms for realising a quantum computer are superconducting qubits. Magnetic fields play an important role in this architecture to tune the frequency of individual qubits, change the coupling between pairs of qubits~\cite{chen2014a} as well as apply parametric drives for quantum limited amplifiers~\cite{yamamoto2008a, mutus2013a} and novel coupling schemes~\cite{chen2014a, caldwell2018}. So called flux bias lines~\cite{dicarlo2009a} are widely used in planar superconducting qubit architectures but so far it has proven to be difficult to implement a similar technology in a 3D architecture~\cite{reshitnyk2016a,kong2015a}. The problems in this architecture arise on one hand due to the difficulty of applying magnetic fields through a superconducting wall and on the other hand achieving fast tunability without compromising the qubit lifetime. %115 words

%inserting them inside a copper cavity could be a solution, but they could couple strongly to the qubits, leading to faster decay times or decoherence [ref?].

%- fast flux tuning possible in planar SC qubit architecure
%- Martinis frequency and coupling
%- difficult for 3D structures or the Leek architecture (?)
%--

%In all experiments involving cavities, injection of a magnetic flux leads to problems. Bias flux lines are widely used in planar superconducting qubit architectures, inserting them inside a copper cavity could be a solution, but they could couple strongly to the qubits, leading to faster decay times or decoherence [ref?].

%By using a copper cavity, it is possible to apply a static magnetic field from the outside without issues, but quick changes of the magnetic field are not possible since eddy currents will prevent the magnetic field to change rapidly, leading to rising times on the scale of several ms, a time that is much larger than the SC qubits lifetime. 

%In cavity QED experiments, superconducting cavities are usually preferred because of their large quality factor, unfortunately the Meissner effect will prevent also static magnetic fields to enter the cavity, so a solution is needed in order to inject a magnetic field inside it. Note that bias lines can't work in this case.

%\subsection*{Our solution}

%- magnetic hose to guide and "focus"

%- maintain fast $\approx$~ns response
In this paper we demonstrate magnetic field control inside a superconducting waveguide cavity using a magnetic hose~\cite{navau2014}. The hose is designed as an effective microwave filter to not compromise the energy relaxation time of the qubit and allow for fast flux control of 3D transmon qubits on time scales \unit[$<$~100]{ns}. This technology can be broadly used to guide magnetic fields from an external coil to a quantum system sensitive to magnetic flux placed inside a metal or superconducting structure or even to mediate magnetic interaction between two quantum systems. %90 words

%- two tunable asymmetric transmon qubits,loop size 200x200 (um), distance 2 mm

%- resonance frequencies qubits: 6.6 Ghz and 6.2 GHz and cavity: 8.1 GHz, anharmonicity: 295 MHz, couplings 5MHz, dispersive shift 4MHz

%V1) The qubits are based on an asymmetric SQUID, with a loop size of 200x200 $\mu m^2$ and an asymmetry factor d = 0.76 (in the design, measured is 0.31), that should have lead to a frequency tunability of around 1 GHz and the maximum frequency was expected to be 7.5 GHz. Our qubit measurements return an asymmetry factor of 0.3, with a tubability of over 3 GHz, and a maximum frequency of 6.6 and 6.2 GHz.

%v2) The qubits are based on an asymmetric SQUID, with a loop size of 200x200 $\mu m^2$. We measured a tunability of more than 3 GHz and a maximum frequency of 6.6 and 6.2 GHz for the qubits. This tunability range is much larger than we designed, at frequencies below four GHz is not possible to use the qubits anymore.

\begin{figure}[t]
   \centering   
    \includegraphics{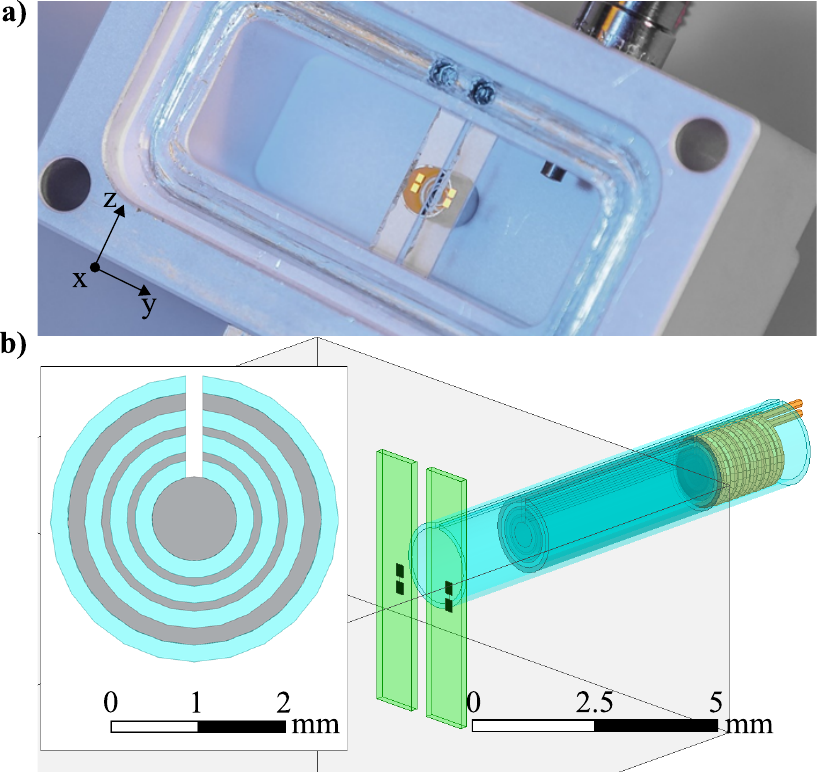} %182, v4 is 161
    \caption{\textbf{(a)} Photograph of one half of a rectangular waveguide cavity with two transmon qubits fabricated on sapphire pieces separated by a distance of about \unit[3]{mm}. The cavity has a hole in the middle of the back wall to attach a magnetic hose. On the back side of the cavity, not visible in the picture, a copper clamp is used to fix and thermalize the hose. \textbf{(b)} Schematic of the setup highlighting the hose, with a coil (yellow) on the external side and a qubit (black structures on the two pieces of sapphire (green)) on the other. Inset: Cross section of the hose, showing the shell structure (grey = ferromagnetic layers, light blue = superconducting layers) with a vertical cut.} %120 words
    \label{Fig1}
\end{figure}
%\textbf{Choosing V2}:

%\textbf{system description} 
Our system, illustrated in Fig.~\ref{Fig1}, is constituted by two transmon qubits separated by a distance of about \unit[3]{mm}, fabricated on two different sapphire pieces. Both sapphire pieces are placed in a rectangular aluminium cavity, which is adapted to the hose by drilling a hole in the middle of one of the faces. %52 words

%\subsection*{magnetic hose}

%- meta material for slow magnetic fields

%- idea from transformation optics

%- $\mu_r^{||} \rightarrow \infty$ and $\mu_r^{\perp} \rightarrow 0$

%- layered structure, cite J

%- so far only DC we show AC

%- field focusing, see Fig.2

\begin{figure}[t]
    \centering
    \includegraphics{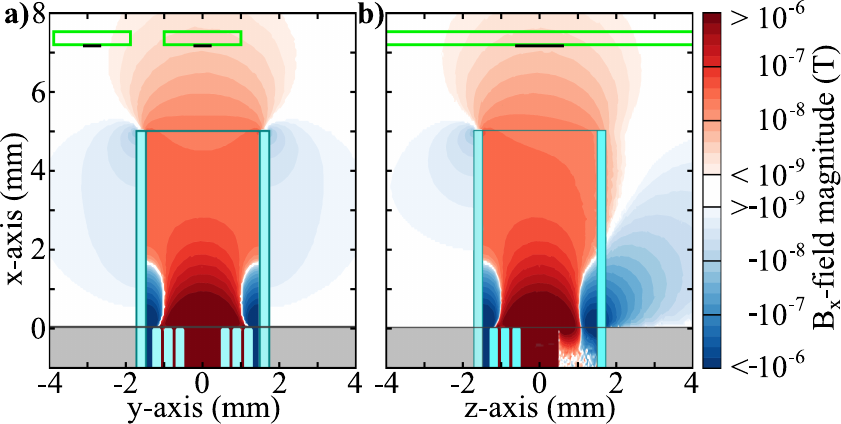} %95 words
    \caption{Top \textbf{(a)} and side \textbf{(b)} view of the setup, showing the magnetic field's x-component on a logarithmic scale. The grey part below ($x<0$) depicts the cavity wall, the thin light blue surfaces indicate the superconducting shells of the magnetic hose, where the magnetic field is zero. The qubits' positions are represented by black lines on green squares, which indicate the sapphire chips. The magnetic field distribution is calculated by a finite element simulation in the case of static magnetic fields. In this case the setup simplifies to its magnetic properties only and the superconductor and ferromagnet are well approximated by materials with $\mu_r=10^{-4}$ and $\mu_r = 10^{4}$, respectively~\cite{sidenote1}.
    The source of magnetic field is a solenoid placed at the entrance of the hose (not visible). Its dimensions and magnetisation correspond to the coil used in the experiment when a current of \unit[10]{mA} is fed. The magnetic field is \unit[3.55]{nT} at the central and \unit[0.75]{nT} at the side qubit.%159
    %, focusing on the magnetic B-field distribution inside and around the hose, notice also the relative qubits' position. \textbf{(b)} Side view of the magnetic B-field distribution.
    } %146 words
    \label{Fig2}
\end{figure}

%\textbf{magnetic hose} 
%A magnetic hose is a device that allows arbitrary routing and transfer of magnetic fields~\cite{navau2014}.
%The magnetic hose is realised by a metamaterial that approximates a fictitious 'ideal material' with magnetic properties of $\mu_r^{||} \rightarrow \infty$ along and $\mu_r^{\perp} \rightarrow 0$ perpendicular to the direction of magnetic field transport~\cite{navau2014}.This 'ideal material' results from calculations using ideas in transformation optics~\cite{Pendry2006,chen_transformation_2010} and is well approximated by a combination of ferromagnetic and superconducting cylindrical layers in alternating order.Whereas the ferromagnetic part has a finite $\mu_r$ in any direction, the superconductor around has \( \mu_r=0 \) and thus confines the magnetic field to the ferromagnetic material and sets the direction of the magnetic field transport. A dense structure of alternating layers is essential for preventing losses and enhancing the transport. 
%
The magnetic hose consists of a set of concentric cylindrical layers. Half of the layers are made of ferromagnetic material (with large magnetic permeability) and the other half of superconducting material (with a permeability effectively $\mu_r\rightarrow 0$). These two types of layers are alternated forming a magnetic metamaterial structure. Such structure exhibits an extremely anisotropic effective magnetic permeability, with very large permeability along the axial direction of the hose, $\mu_r^{||} \rightarrow \infty$, and a very small permeability in the (radial) perpendicular direction, $\mu_r^{\perp} \rightarrow 0$. Intuitively, the ferromagnetic layers provide large permeability components in all directions, but the interlayered superconducting layers shield any radial magnetic field component. The design of this metamaterial structure~\cite{navau2014}, which results from concepts in transformation optics~\cite{Pendry2006,chen2010a}, allows to transfer static magnetic fields over long distances. To that effect, a dense structure of alternating layers is essential for preventing losses and enhancing the transport. %147 words

%Following this idea, we built a hose consisting  of four superconductor+ferromagnetic layers, with mu-metal as ferromagnet and aluminium as superconductor (Fig \ref{Fig1}b). The three innermost layers of the hose are \unit[10]{mm} long and enclosed by an elongated aluminium shell of \unit[20]{mm} length. The elongated outermost shell pushes the magnetic field further into the cavity. Additionally to the elongated outermost shell we introduce a cut through all layers along the magnetic hose in comparison to the original design \cite{navau2014}.
%All aluminium shells are \unit[200]{$\mu$m}, the two inner most mu-metal shells are \unit[100]{$\mu$m} and the outermost mu-metal shell is \unit[200]{$\mu$m} thick. All shells are wrapped around a \unit[1]{mm} thick mu-metal wire. The coil at one end of the hose has 10 turns, a diameter of \unit[3]{mm}, and a length of \unit[4]{mm}, resulting in a self inductance of \unit[126]{nH} in free space. When the coil is attached to the hose, the self inductance is reduced to \unit[81]{nH} due to the outermost superconducting shell surrounding it. 
%
Following this idea, we built a hose consisting of four ferromagnetic layers made of mu-metal and four superconducting layers made of aluminum.
The seven innermost layers are \unit[10]{mm} long and the outermost aluminium shell is \unit[20]{mm} long. This external superconducting shell pushes the magnetic field further into the cavity. Additionally, we introduce a cut through all layers along the magnetic hose (see Fig~\ref{Fig1}b).
All aluminium shells have a thickness of \unit[200]{$\mu$m}, the two innermost mu-metal shells are \unit[100]{$\mu$m} thick and the outermost mu-metal shell is \unit[200]{$\mu$m} thick. All shells are wrapped around a \unit[1]{mm} thick central mu-metal wire.
The coil at one end of the hose has 10 turns, a diameter of \unit[3]{mm}, and a length of \unit[4]{mm}, resulting in a self inductance of \unit[126]{nH} in free space.
When the coil is attached to the hose, the self inductance is reduced to \unit[81]{nH} due to the outermost superconducting shell surrounding it. %140 words

The question arises if one can use such a magnetic hose for transporting a magnetic flux pulse generated outside to the inside of a superconducting cavity and locally act on a flux sensitive qubit. %alternatively: in the cavity center, but 2 times the word "inside"sounds bad
A magneto-static simulation shows that the magnetic hose is able to route magnetic field through a hole into a superconducting box as depicted in Fig.~\ref{Fig2}.
At first glance this seems to contradict the concept of flux quantization, which prevents any net magnetic flux from passing through the area of a closed superconducting ring that has been cooled in absence of magnetic field.
However, the magnetic field is not only guided into the cavity, it is also routed back to the outside through the same hole.
As a result, the total magnetic flux through the hole is always zero and flux quantization is fulfilled. %137
%Flux quantization has to be taken into account also for the superconducting layers of the hose itself.
%Implications of flux quantization also extend to the superconducting layers forming the hose itself. For this reason, a magnetic hose can transport magnetic field only if all layers along the magnetic hose are cut, as shown in the inset of Fig.~\ref{Fig1}b. This cut prevents super or eddy currents from flowing in closed circles to counteract the magnetic field from being transported through. As a consequence not only static but even fast alternating magnetic fields can be transported.
Implications of flux quantization also extend to the superconducting layers forming the hose itself. For this reason, a magnetic hose can transport magnetic field only if all superconducting layers are cut along their length, preventing super-currents to flow in closed circles. Since ferromagnetic layers also have some electrical conductivity, the cut is also extended to them in order to minimize circular eddy currents (see the inset of Fig.~\ref{Fig1}b). As a result, not only static but even fast alternating magnetic fields can be transported. %83 words

Despite the magnetic hose opening a link for magnetic fields through the cavity wall, it is designed to prevent electromagnetic waves inside the cavity from leaking out to the environment. The elongated part of the outermost shell acts as a $\lambda/4$ resonator with a resonance frequency around \unit[10]{GHz}. Any electro-magnetic wave below this resonance is reflected and does not reach the ferromagnetic core of the hose. The cut introduced in the hose layers acts as a rectangular waveguide with a cut-off frequency of more than \unit[100]{Ghz}. Thus, the magnetic hose does not provide any additional loss channel for microwaves at the qubit and cavity frequencies, which was verified by finite element simulations. In summary, a magnetic hose acts as a filter between the coil and the qubit and, due to its shape, the transported magnetic field is applied very locally. %141 words

%Thus, the qubit can be placed further away from the ferromagnetic part of the hose without losing much field strength along the distance.
%A qubit close to a ferromagnet might couple inductively to it and is affected by noise from the ferromagnet.

%[better if Stefan writes this section]
%The magnetic field transport can be improved by small changes of the current setup. As suggested in ref 'hose paper' a higher density of ferromagnetic and superconductor layers improves the performance. A very dense structure could be reached with sputtering. Since the magnetic field transport through the magnetic hose structure is lossy, one can shorten the hose to decrease losses during the transport. Further improvement is gained by elongating the inner superconducting shells to the length of the outermost shell. As a result magnetic field lines are pushed further.

%- apply DC current up to xx mA
%- apply voltage using the AWG, around 3.3V are necessary for half quantum jump for qubit1, over 15V for qubit 2, the map was done using -+1V, and the top part is stiched from another cooldown.
%- see Fig.3

\begin{figure}[t]
\centering
    \includegraphics{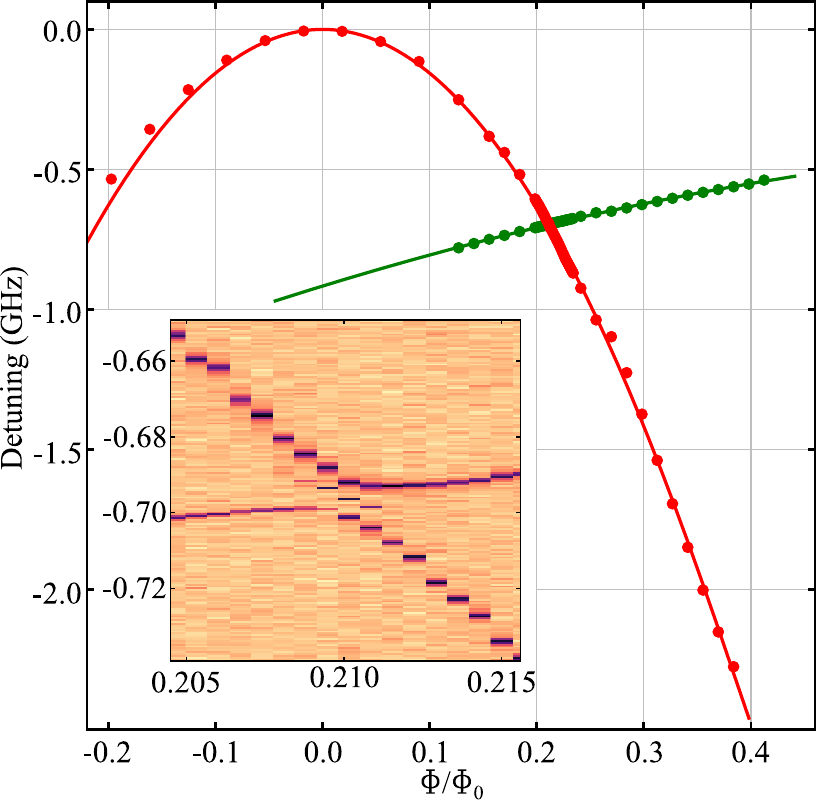} %167, v3 is 132
    \caption{Detuning from the maximum transition frequency of both qubits as a function of the applied magnetic flux (data combined from two independent cool-downs): the plot shows that the qubit in the centre (red dots) is more tunable than the one on the side (green dots). A fit~\cite{koch2007} to the data is plotted in solid red and green lines. From this fit we extracted the asymmetry of the SQUID loop and the maximal resonance frequency. A zoom of the avoided crossing is shown in the inset, with a splitting of $2g \approx \unit[10]{MHz}$. The line intersecting the avoided crossing, is the 2 photon transition to the state where both qubits are excited.}%106
    \label{Fig3}
\end{figure}

%- tune qubit over about 1 Flux quant

%- focusing qorks ratio about 1:5??

%- stable enough to tune into avoided crossing

%\textbf{measurement DC flux} %The first part of the experiment aims to check frequency tunability by applying a DC voltage to the coil, provided by the AWG, as described above.
After assessing the functionality of the hose through numerical simulations, we prepared the experiment~\cite{supplementary} with the aim of verifying the qubit resonance frequency tunability by applying a DC voltage to the coil. In this experiment we use an Arbitrary Waveform Generator (AWG) channel to provide the above mentioned coil excitation. The flux map for both transmon qubits is shown in Fig.~\ref{Fig3}. The spectroscopy is performed using a saturation pulse with variable frequency, followed by a readout pulse at the cavity resonance frequency, for each value of the AWG output voltage. The maximum applied magnetic field is limited by the coil size and the maximum output voltage of the AWG. The measurement shows that the qubits are tunable as intended. The qubit resonance frequencies are extracted from the measurements and fitted. The flux map also demonstrates the local control capabilities of the hose, as one qubit is more affected by the injected magnetic flux than the other. From the fit we can extract that the hose is around 5 times more effective on the centre qubit, which agrees well with numerical simulations. This factor can be further improved by using an optimised smaller hose design with a funnel-shaped tip and placing the qubits closer to the hose (see Fig.~\ref{Fig2}). %208 words

%\textbf{[optional-start, just an idea]}
%From the measurement we noticed that the qubits trapped some magnetic flux, most probably due to a slight magnetisation of the hose's core. Since the external magnetic field was limited by the coil size and the AWG output voltage, we extended the measurement in a second experiment by re-trapping flux. The two parts are plotted together in the figure, the fit (continuous line) doesn't agree well with the data points because of a slight change in the Josephson energy and the asymmetry factor due to re-cooling. \textbf{[end-optional]}

The data points density has been increased in the avoided crossing region to measure the qubit-qubit interaction, which is \unit[$\approx$ 5]{MHz}. The weak interaction between the qubits is expected from finite element simulations due to the relative qubits distance and position inside the cavity~\cite{dalmonte2015b}. %44 words

%- to figure out fast flux capabilities we use sequence shown in Fig 4.a

%- limited by pulse, compromise between frequency and timing resolution

%- fast rise time and stable over time

%\textbf{measurement fast} 

\begin{figure}[t]
    \centering
    \includegraphics{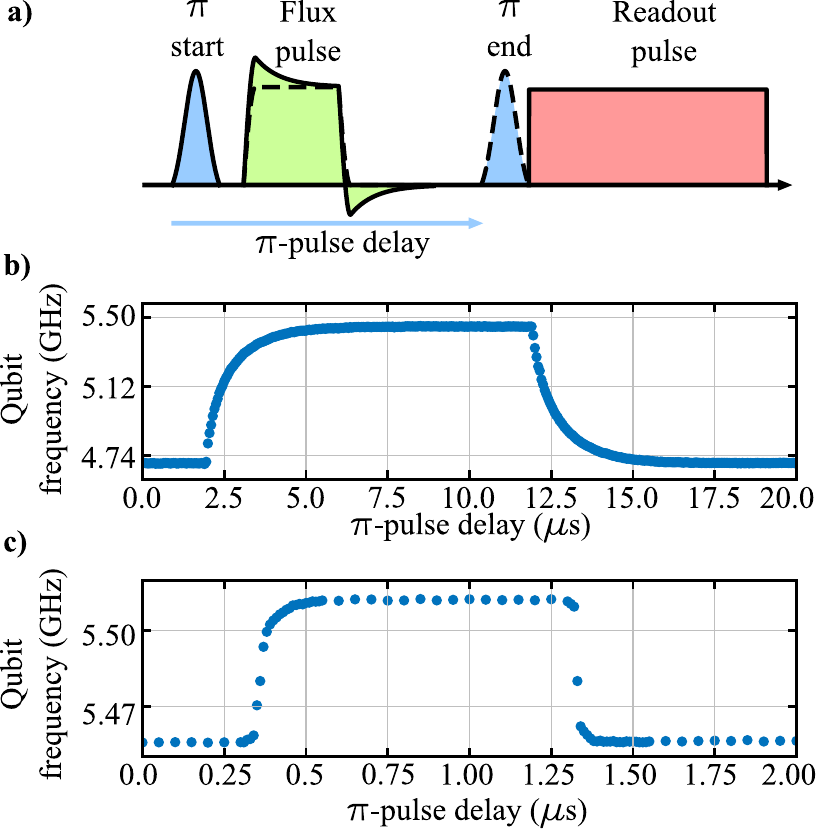} %172 words
    \caption{Measurements of the qubit tunability with AC magnetic field. \textbf{(a)} Pulse sequence used to perform the fast flux measurements: the Gaussian pulse (blues) is a $\pi$-pulse used to excite the qubit; the flux pulse is shown in green, note that the dashed flux-pulse line is used to get the measurement (b) whereas the continuous line flux pulse is used to get the measurement (c); finally the red pulse is used to readout the qubit state. All the pulses but the $\pi$-pulse are fixed in time, while the frequency and delay of the $\pi$-pulse are linearly changed. The initial position of the pulse defines our time-axis origin and it stops just before the readout pulse, avoiding an overlap. \textbf{(b)} This measurement shows a qubit resonance frequency variation of around \unit[700]{MHz}. The measured exponential rise time is about \unit[2]{$\mu$s}. \textbf{(c)} Using the AWG the flux pulse is modified to have an initial overshoot, in order to speed up the rise time that is found in measurement (b). The step size on the time-axis is \unit[20]{ns} but the resolution given by the $\pi$-pulse is around \unit[50]{ns}.} %186
    \label{Fig4}
\end{figure}

In order to assess the fast flux capabilities of our system, we perform another experiment using the sequence shown in Fig.~\ref{Fig4}a. To excite the qubit we use a Gaussian amplitude modulated $\pi$-pulse, with a standard deviation of around \unit[70]{ns} and a total length of \unit[400]{ns}, generated by the AWG and up-mixed with an IQ-mixer. The $\pi$-pulse length is a trade off between time and frequency resolution for this particular experiment. Another AWG channel is again used to excite the coil, but with pulses added to an optional DC offset. During the measurements, the flux-pulse and the readout pulse have a fixed delay. The frequency and the delay of the $\pi$-pulse are changed linearly, to track the qubit resonance frequency during the flux pulse. %130 words

The first measurement (Fig.~\ref{Fig4}b) is obtained using a square flux-pulse for the coil excitation (dashed line in Fig~\ref{Fig4}a). In order to avoid ringing effects caused by the AWG, the square pulse's edges are smoothed by a $\pm\sin(\frac{\pi}{2}\frac{t}{\tau})$ rise or fall for \unit[20]{ns}, where $\tau=\unit[20]{ns}$.
%is a composition of the first quarter of a sinusoidal and a step function, with a total slope of $\unit[20]{ns}$.
The measurement result shows a smooth change of the qubit resonance frequency, with an exponential rise time on the order of \unit[2]{$\mu$s}. The source of this rise time is still under investigation and is probably caused by eddy currents in the copper clamp holding the hose. A slightly modified clamp design can easily mitigate this effect and will be investigated in future experiments. %107

In a second set of measurements we speed up the system response through a simple pulse shape, adding an overshoot and an exponential decay (continuous line in Fig.~\ref{Fig4}b) to the flux pulse. Obtaining the result shown in Fig.~\ref{Fig4}c, it is evident that now the change in the qubit resonance frequency occurs on time scales $<$~\unit[100]{ns}. The frequency of the qubit remains stable after around \unit[300]{ns} without any ringing being evident. The maximum flux step is limited by the maximum available AWG voltage to create the overshoot. A further speed up could be achieved by a proper linear system de-convolution. %101

%- T1 and T2 as a function of frequency, T1 is 12 to 25 us and T2 is 600 ns to 2.2us, with TEcho 8x times larger than T2 

%- cavity Q change from XX to 25k, cold

%- qubit population: 3% approx 80mK, check ref
% anharmonicity increased from 265 MHz to 295 MHz

%\textbf{coherence properties with hose} 
After checking the hose capabilities to apply a magnetic field, we estimate the effect of the hose on the system by measuring $T_1$, $T_2^*$, qubit population and the internal quality factor of the cavity. %34 words
The $T_1$ of our qubits is Purcell limited and varies from \unit[15]{$\mu$s} at the sweet spot to \unit[30]{$\mu$s} down at \unit[4]{GHz}. The $T_2^*$ varies from about \unit[2.2]{$\mu$s} at the sweet spot, down to \unit[600]{ns} far from it, with $T_{2E}$ being eight times larger than $T_2^*$ independent of flux bias. To investigate possible slow frequency noise introduced by the hose, we decided to perform experiments without the hose. For these measurements, we close the hole in the cavity wall and  find comparable values for $T_2^*$ and $T_{2E}$. We believe that the slow frequency noise in the experiments stems from the large SQUID loop and associated flux noise. It is even further aggravated by using the big capacitive shunt pads to close the SQUID loop at the top and bottom~\cite{stan2004,ithier2005}. %133 words

The qubit population has been estimated using the method shown in~\cite{geerlings2013a}. The result obtained is \unit[80]{mK}, comparable to the result obtained in the same reference and reference~\cite{jin2015e}. This result is very typical for our setup and compatible with other experiments done in our lab that do not use hoses. %51 words

In order to measure the internal quality factor $Q_{\rm{int}}$ of the cavity with the hose, a reflection measurement using a microwave circulator is done. The measured $Q_{\rm{int}}$ is about $25\cdot10^3$, much less than the expected value in the range of $10^5 - 10^6$. Simulations show that this reduction in $Q_{\rm{int}}$ is caused by the fundamental mode's surface currents being dissipated when they cross the gap between the cavity wall and the hose's outermost shell~\cite{brecht2015a}. We experimentally verified this by sealing the gap with indium and measuring again $Q_{\rm{int}}$, reaching $2.75\cdot10^5$. To completely circumvent any lossy seam by introducing the magnetic hose, we suggest to fabricate the outermost shell as part of the cavity when it is machined. %116 words

%\textbf{future steps}

This first magnetic hose design can be improved by a factor of 10 to transfer higher magnetic fields and to act more locally at the qubit's position~\cite{supplementary}. Magnetic field transport is enhanced by increasing the layer density, shortening the total length, and elongating all superconducting shells. Addressing can be simply improved by reducing the radius or by forming a funnel-like structure similar to field concentrators~\cite{prat-camps2013,zhao2018a}.
%Both come at the expense of an overall reduced performance as there is less metamaterial cross-section in a smaller hose.
Moreover, it is worth to mention that an optimally designed coil increases the transferred magnetic field. %78 words

%The maximal amount of magnetic field being transported through the hose is mainly limited by the ferromagnet's saturation. This is irrelevant for our setup, however for other applications it is worth to mention that hundreds of \unit{mT} can be transported depending on individual ferromagnetic materials~\cite{magweb}. The critical field of the superconducting shells are not expected to limit the transport, since the radial magnetic field component is much smaller than the axial one. Furthermore high-Tc superconductors like YBCO are able to shield magnetic fields of \unit[$>$1]{T} completely~\cite{wera_2017}.

The maximal amount of magnetic field being transferred through the hose is limited by the used ferromagnetic and superconducting materials.
%The overall performance of the hose will depend on the particular values of field one intends to transfer and the used ferromagnetic and superconducting material.
In our setup, fields on the order of $<\unit[10^{-6}]{T}$ are sufficient to control the qubits and, thus, materials can be assumed to deviate very little from their ideal linear behaviour. The transfer of much higher fields would induce non-linear responses of the materials, which one would need to analyze carefully. On one hand, the superconducting shells are not expected to significantly limit the hose performance, since radial components of magnetic field (the ones that the superconductor shields) are much smaller than axial components. Furthermore, high-Tc superconductors like YBCO are able to shield magnetic fields of \unit[$>$1]{T} completely~\cite{wera2017}. Ferromagnetic shells, on the other hand, are expected to be the main limiting factor for high-fields transfer. Numerical calculations indicate, though, that fields on the order of hundreds of \unit{mT} can be transported through hoses made of high-field-saturation ferromagnetic materials like soft irons or electrical steels~\cite{noauthor_magweb_nodate}. %163

%The transfer of much higher magnetic fields would be mainly limited by the non-linear behaviour of the ferromagnet, in particular, by its saturation. 
%This is irrelevant for our setup, however for other applications it is worth to meantion that hundreds of \unit{mT} can be transported depending on individual ferromagnetic materials~\cite{magweb}. The critical field of the superconducting shells are not expected to limit the transport, since the radial magnetic field component is much smaller than the axial one. Furthermore high-Tc superconductors like YBCO are able to shield magnetic fields of \unit[$>$1]{T} completely~\cite{wera_2017}.

%improvements of the hose:
%x reduce radius of the hose to act more local on a single qubit.
%x elongate all superconducting shells, pushes the magnetic field better into the cavity.
%x shorter core, to reduce losses inside the hose
%x thinner and denser layer structure, feasible with sputtering, comes closer to the 'ideal material' and increases performance
%x smaller coils, that are closer to the magnetic hose's core. 

%From simulations we know that the hose can be further improved by reducing its dimensions and geometry, with the improved design we should be able to focus the magnetic field further to a specific qubit.

%Cavity Q should be improved with thinner hose - move to wavguide and MSR

%\textbf{What else can we use it for}
The experimental work presented in this paper shows the potential of using a magnetic hose to control various quantum systems. The hose is most applicable in scenario's where a local applied magnetic field is required, combined with the ability to generate the magnetic field at a distant location and route it to the quantum systems of interest. Most obviously, it would be ideally suited to individually control qubits in superconducting circuits architectures like~\cite{zoepfl2017,rahamim2017a,axline2016} without compromising coherence times. Moreover, due to the ability to apply time varying signals, one could also use the hose to drive parametric amplifiers~\cite{mutus2013a,Mutus2014} or tunable couplers~\cite{chen2014a}. The magnetic hose could also be an ideal candidate to introduce time varying magnetic fields in a hybrid architecture e.g. Rydberg atoms coupled to superconducting microwave cavities~\cite{stammeier2017c}. The hose might also be applicable to control the magnetic field in gate defined double quantum dots~\cite{Petta2005, Stockklauser2017a} or even hybrid scenarios where the quantum dot is coupled to a transmon qubit \cite{scarlino2018}. It could also allow new designs of NMR sensors based on NV centers~\cite{arai2015,glenn2018} or even enable magnetic coupling between NV centers~\cite{gaebel2006}.

%\textbf{Summary}
In summary we have shown that a magnetic hose can be used to change the magnetic flux inside an aluminum waveguide cavity in less than \unit[100]{ns}. The magnetic hose is constructed such that it does no compromise the energy relaxation time of the qubit on the several \unit[10]{$\mu$s} timescale. Furthermore, the hose can be used to localise the magnetic field and address predominantly one of the qubits inside the cavity. Future improvements to the hose design promise to increase the field transport by a factor of 10 and localize the field to an even smaller volume. As such, the hose is ideally suited to implement fast flux bias lines in 3D~\cite{axline2016} and waveguide~\cite{zoepfl2017} circuit QED architectures. %118 words

%\section{Acknowledgement} 
We gratefully acknowledge fruitful discussions with O. Romero-Isart and thank M.L. Juan and A. Sharafiev for feedback on the paper. Facilities use was supported by the KIT Nanostructure Service Laboratory (NSL). OG and GK are supported by the Austrian Federal Ministry of Education, Science and Research (BMWF), SO und MZ are funded by the European Research  Council  (ERC)  under  the  European  Union’s Horizon  2020  research  and  innovation  program  (grant agreement n$^{\circ}$ 714235).  MZ is also supported by the Austrian Science Fund FWF within the DK-ALM (W1259-N27). JPC is funded by the European Research Council (ERC-2013-StG 335489 QSuperMag) and the Austrian Federal Ministry of Education, Science and Research (BMWF).

\newcommand{\D}{\Delta}
\newcommand{\tD}{\tilde{\Delta}}
\newcommand{\K}{K_{PP}}
\newcommand{\bn}{\bar{n}_P}
\newcommand{\G}{\Gamma}
\newcommand{\LH}{\underset{L}{H}}
\newcommand{\HL}{\underset{H}{L}}
\newcommand{\blue}[1]{{\color{blue} {#1}}}

%\bibliography{references_main}
%

\clearpage

\onecolumngrid
\begin{center}
\textbf{\large Supplementary Material}% Force line breaks with \\
\vspace{5mm}
\end{center}
\twocolumngrid

\setcounter{equation}{0}
\setcounter{figure}{0}
\setcounter{table}{0}
\setcounter{section}{0}
\setcounter{page}{1}
\makeatletter
\renewcommand{\theequation}{S\arabic{equation}}
\renewcommand{\thefigure}{S\arabic{figure}}
\renewcommand{\bibnumfmt}[1]{[S#1]}
\renewcommand{\citenumfont}[1]{S#1}

\section{Setup and measurements}
Both transmon qubits are based on an asymmetric SQUID ($d=0.31$)~\cite{koch2007s}, with a loop size of \unit[200~x~200]{$\mu \mathrm{m}^2$} and an anharmonicity of \unit[295]{MHz}. We measured a tunability of more than \unit[3]{GHz} and maximum frequencies of \unit[6.6]{GHz} and \unit[6.2]{GHz} respectively. Below \unit[4]{GHz} the dispersive shift, the $E_J/E_C$ ratio and anharmonicity are drastically reduced and it becomes difficult to probe the qubits. 

The qubits are capacitively coupled to the first mode of the cavity, with a resonance frequency of 8.102 GHz. The coupling is such that the dispersive shift $\chi \approx $~\unit[5]{MHz} at the qubit's maximum transition frequency. The total quality factor of the cavity is limited by the output coupling, engineered to have $\kappa \approx \chi$ at the maximum qubit frequency. %62 words

A cryostat input line with a total of \unit[50]{dB} attenuation and DC~-~\unit[12]{GHz} filtering is used to couple to the cavity. A similar line, with only \unit[20]{dB} of attenuation at the \unit[4]{K} plate and DC~-~\unit[80]{MHz} filtering, is used for driving the coil of the magnetic hose. Additional filtering and better line thermalization is provided by custom-made \unit[50]{$\Omega$} absorptive coaxial filters. %63 words

To perform a readout, a square, finite-length microwave pulse at the cavity frequency is generated and sent to the system. The signal, transmitted through the cavity, undergoes additional filtering and passes through two isolators, it is then amplified by a HEMT at 4K and by a low noise classical amplifier at room temperature. After that, the signal is down-mixed and digitally recorded by an ADC. Qubit rotations are performed by Gaussian amplitude-modulated pulses generated by an Arbitray Waveform Generator (AWG) and up-converted from around \unit[200]{MHz} to the qubit frequency. Another AWG channel is used to provide pulsed and DC voltage to the coil, allowing us to reshape the pulses used for the experiment, like described in the main paper.

\begin{figure}[t]
    \centering   
    \includegraphics{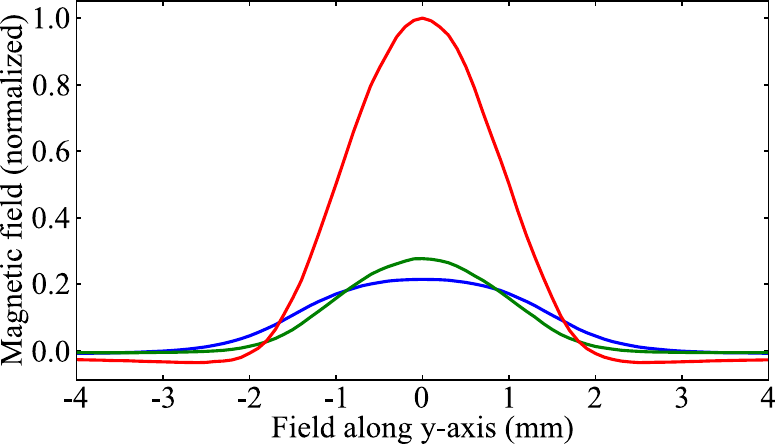}
    \caption{Simulated magnetic field intensity in x-direction along the y-axis at a distance of \unit[1]{mm} from the hose for three different hose and coil designs attached to a closed superconducting cavity. The blue line represents the hose used in the paper. The green line considers a hose with higher layer density and elongated shells in combination with a big coil which outer diameter matches the outermost superconducting shell inner diameter. Finally, in red, the maximum improvement is achieved by optimizing both the hose and the coil that magnetizes the ferromagnetic layers in the optimal way.}
    \label{FigS1}
\end{figure}

\section{Enhancing the magnetic field transport}

% hose parameters transport:
    % - layer density
    % - length
    % - elongating all superconducting shells
% hose parameters field distribution:
    % - smaller
    % - funnel shape
% external parameters
    % - coil
    % - environment, remove shielding material from coil
    
The magnetic hose presented in the main paper proofs the ability of guiding a magnetic field from the outside to the inside of a superconducting cavity.
This first design, however, can be improved to enhance the field transport and to address the magnetic field more locally.

First, the field transport will be enhanced by shortening the total length, increasing the layer density, and elongating all superconducting shells of the magnetic hose on the output side.
The length of the inner core can be decreased to minimise the amount of magnetic field leaking and turning back at the cut before entering the cavity.
According to Ref.~\cite{navau2014s} a higher density of ferromagnetic and superconducting layers increases the magnetic field transport as the metamaterial approaches the ideal material resulting from their calculations.
Large densities (layer thickness $\approx \unit[100]{nm}$) and small hose sizes ($\approx \unit[100]{\mu m}$) could be reached with sputtering methods starting from a central ferromagnetic core.
Besides the outermost superconducting shell, elongating all superconducting shells will push the magnetic field further into the cavity, since the magnetic field is forced to pass the shells before being routed back. Obviously, elongating the ferromagnetic shells will also enhance the field transport.
However, we prefer to have a decent distance between our qubits and the ferromagnet to avoid additional losses for the qubit.   

Second, the field distribution at the output side is controlled by the shape of the magnetic hose.
By simply reducing the radius of the magnetic hose, the transferred magnetic field is addressed more locally. 
The addressability is further increased by forming a funnel shaped tip, similar to magnetic field concentrators~\cite{prat-camps2014s}.
In such a configuration one could transfer and concentrate the field from a large magnetic source at the input side to a tiny spot at the output side of the hose.

Lastly, the design of the coil at the input side of the hose can be optimized to increase the magnetic field transport and to address the magnetic field more locally.
It is essential to bring as many turns as possible very close to the metamaterial at the input side of the magnetic hose in order to maximize the magnetic flux injected to the inner ferromagnetic shells.
As shown in the simulations, to get magnetic field into the cavity it is necessary that the outer ferromagnetic shells are magnetized in the opposite direction, guiding the magnetic field lines back to the coil.
Since the magnetic field strength decreases over the distance, it is more favourable to attach a discoidal coil to the hose instead of a cylindrical coil.
Furthermore one should reduce any material counteracting the generated magnetic field to a minimum in the vicinity of the coil.
The combined design of the magnetic hose structure and the attached coil is essential for reaching the intended magnetic field distribution at the output side (Fig.~\ref{FigS1}).

\section{Shielding properties}

% shielding from external magnetic fields
    % - elongated outermost shell, shields entrance of magnetic hose from surrounding magnetic fields
    % - only fields close to the metamaterial will be transported
% shielding from external µ-wave fields
    % - outermost shell protects coil from picking up external fields, due to circular waveguide
    % - µ-waves cannot enter hose from outside, due to circular waveguide structure or lambda/4 filter.

Our experiment is surrounded by a copper and a cryoperm shield, providing the usual basic shielding as used in similar experiments~\cite{kreikebaum_optimization_2016s}.
The bulk aluminum cavity in the superconducting state shields its inside perfectly from external applied magnetic fields, due to the Meissner effect.
Introducing a magnetic hose to the cavity opens a link not only for bias fields but also for any external magnetic fields.
However, the elongated outermost superconducting shell at the input side provides sufficient shielding from unwanted external magnetic fields.
In our experiment we do not see any shift of the flux map recorded on different days within the linewidth of the qubit.
Thus, we state sufficient shielding from external magnetic fields and no magnetization of the ferromagnetic material.
The coil inside the elongated outermost shell is the only contributor to any frequency shift of both qubits in our experiment.
To remove any doubts, one could build a proper superconducting housing at the input side to ensure perfect isolation from external magnetic fields below the critical field of the shield.

%This also means, that an external, undesired, field will not be able to couple through the hose into the cavity and thus the shielding effect of the cavity is not compromised.  indicates also low sensitivity to external magnetic fields.

In the microwave domain any material with high conductivity reflects microwaves.
Thus, microwave shielding is provided fairly easily for instance by a sheet of copper.
Since the magnetic hose is a combination of two highly conductive materials, microwaves are expected to be reflected from it.
However, one has to introduce a cut to the magnetic hose to guide fast magnetic flux pulses through it.
This cut turns out to act as a rectangular waveguide, providing a channel for microwaves to pass through.
Due to the height and width of \unit[1.2]{mm} and \unit[0.4]{mm}, respectively, the cut-off frequency is above \unit[100]{GHz} and provides more than \unit[20]{dB/mm} of attenuation for frequencies below \unit[60]{GHz}.
Similarly, the elongated outermost shell at the input side protects the embedded coil from microwave radiation that could induce a voltage drop leading to flux noise.
The outermost shell has an inner diameter of \unit[3]{mm} and acts as a circular waveguide with a cut-off above~\unit[50]{GHz} resulting in an attenuation larger than \unit[10]{dB/mm} for frequencies below \unit[20]{GHz}.

\begin{figure}[t]
    \centering   
    \includegraphics{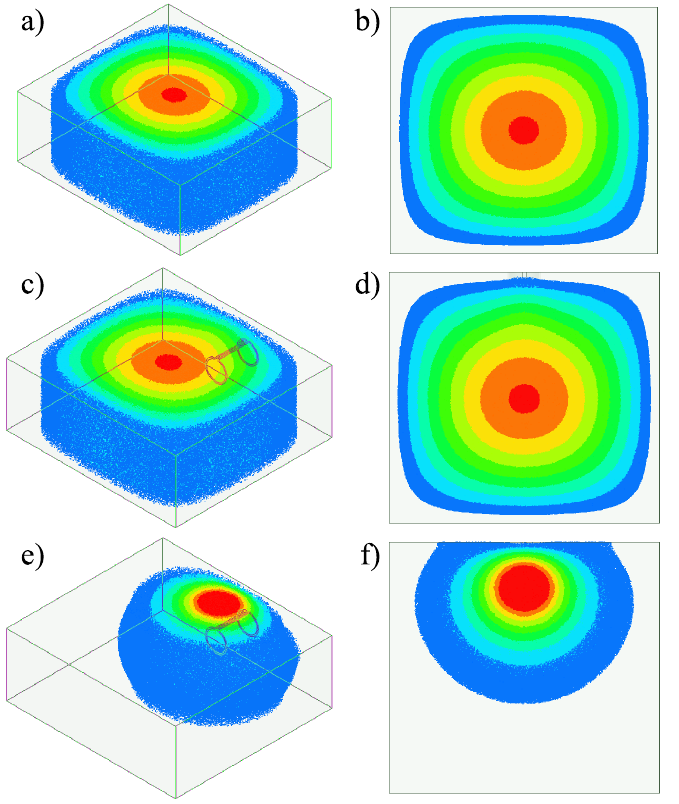}
    \caption{First mode of the bare cavity (\unit[8.17]{GHz}) in isometric \textbf{(a)} and top \textbf{(b)} view. First mode of the cavity with hose (\unit[8.23]{GHz})
    in isometric \textbf{(c)} and top \textbf{(d)} view. Fundamental hose mode (\unit[10.46]{GHz}) of the cavity with hose in isometric \textbf{(e)} and top \textbf{(f)} view. The colormap depicts the magnitude of the electric field  qualitatively, ranging from blue (minimum) to red (maximum).}
    \label{FigS2}
\end{figure}

\section{Effect on cavity modes}

% fundamental mode
    % - shifted to higher frequencies
    % - field maximum is slightly shifted towards hose
    % - surface current distribution is modified
    % - reducing Qint
% hose mode
    % - lambda/4 resonator
    % - depending on length

Introducing the magnetic hose into the cavity reduces its volume and hence shifts its fundamental mode to a higher frequency compared to a bare cavity of the same size.
In addition, the electromagnetic field becomes distorted in the vicinity of the hose to match the condition of zero tangential electrical field along the outermost shell.
For that reason the induced currents of a mode inside the cavity do not only flow on the cavity walls, they also have to flow on the outermost shell.
The outermost shell effectively becomes part of the cavity and therefore has to be well connected to the cavity wall.
Otherwise, currents crossing the gap between the cavity wall and the outermost shell dissipate and limit the internal quality factor of any cavity mode.
In that sense one has to consider the orientation of the cut, pointing always in the direction of the electrical field component of the mode of interest.
%Inside the outermost shell no currents are induced at the fundamental frequency.
The hose itself acts as a $\lambda/4$-resonator inside the cavity and hence it introduces additional modes depending on its length only, independent of the cavity volume.

In the case of our setup, the first three modes of the bare cavity are at \unit[8.17]{GHz}, \unit[12.62]{GHz}. and \unit[13.21]{GHz} according to finite element simulations.
Adding the magnetic hose to the simulations has two effects on the modes inside the cavity.
First, in comparison to the bare cavity, a new mode at \unit[10.46]{GHz} appears.
This mode is dominated by the dimensions of the magnetic hose.
Second, the cavity-dominated modes are slightly shifted to \unit[8.23]{GHz}, \unit[12.63]{GHz}. and \unit[13.20]{GHz}.
The field distribution of selected modes are illustrated in Fig.~\ref{FigS2}.

\begin{figure}[t]
    \centering   
    \includegraphics{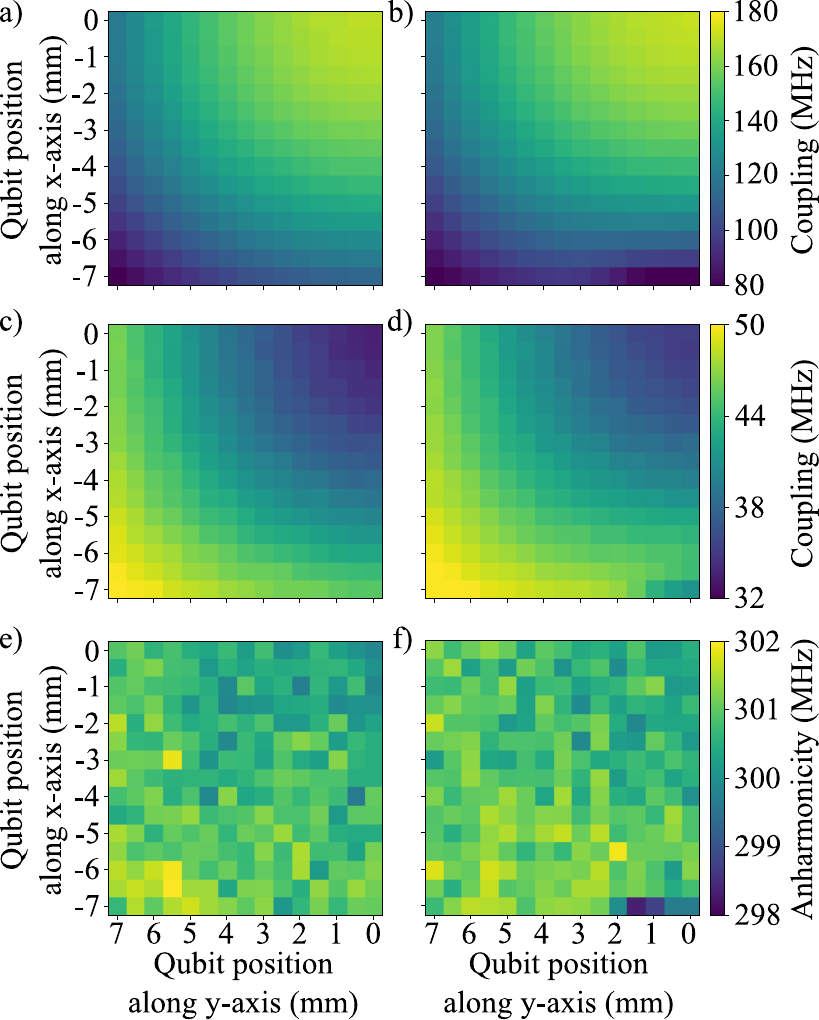}
    \caption{Coupling between transmon and cavity in a bare cavity \textbf{(a)} and a cavity with a hose \textbf{(b)}. Coupling between two transmons in a bare cavity \textbf{(c)} and a cavity with a hose \textbf{(d)}. Anharmonicity of a transmon in a bare cavity \textbf{(e)} and in a cavity with a hose \textbf{(f)}. The plots result from finite element simulations. The transmon is moved in half millimeter steps through one quarter of the cavity, where the origin is set at the centre of the cavity. The coupling between two transmons is evaluated by moving both side by side in half millimeter steps, keeping a constant distance of \unit[2]{mm} between them. The fluctuations of the anharmonicity in \textbf{(e)} and \textbf{(f)} stem from the finite numerical accuracy of the simulations.}
    \label{FigS3}
\end{figure}

\section{Effect on transmons}
% main changes:
    % X - change of E_c
    % X - change of coupling between qubit and cavity
% main consequences:
    % X - interestingly E_c varies only very little ~1%
        % X --> thus frequency and anharmonicity are changed very little
        % X --> maybe more important: transmon stays in transmon regime 
    % - more interestingly the coupling to the cavity
        % X --> depends on the distance between hose and qubit
        % X --> increases a little when qubit is shifted to the side
        % --> independent of the length of the hose inside the cavity, as long the distance between qubit and the hose is constant
    % x - coupling between two qubits
    % x - coupling between hose and qubit

The presence of the magnetic hose does not have any crucial influence on the parameters of the transmon.
In the vicinity of \unit[$\approx$~0.5]{mm} close to the edge of the outermost shell, the capacitive energy of the transmon decreases by \unit[$\approx$~1]{\%} maximally.
Thus, its frequency and anharmonicity are affected by the same (negligible) amount.     
The coupling between the transmon and the cavity is maximally decreased from \unit[113]{MHz} to \unit[78]{MHz}, when the transmon is put \unit[0.5]{mm} close centrally in front of the hose.
This is caused by the local deformation of the electromagnetic field in the vicinity of the hose.
One can compensate the decreased coupling to the cavity by increasing the total length of the transmon.
The coupling between two transmons decreases only if both are closer than \unit[1]{mm} to the hose in comparison to the bare cavity.
Again, this is caused by the local distortion of the electromagnetic field of the transmons and can be compensated by decreasing the distance between them or increasing the pad size.
The coupling between the hose mode and the qubit is essentially zero due to symmetry.

%\bibliography{references_supp}

\begin{thebibliography}{37}
\bibitem {sidenote1}When a holed superconductor is modelled as
  a material with a relative permeability $\mu_r\rightarrow0$, no conditions on
  the magnetic flux threading the hole are imposed. In our calculations, the
  zero-flux condition through the hose is imposed by considering the whole
  closed superconducting cavity. The cavity walls shield the magnetic field in
  such a way that the total magnetic flux crossing the hole is forced to be
  zero to fulfil $\nabla \cdot \mathbf{B}=0$ everywhere in space.
\end{thebibliography}
%

\end{document}